\documentclass[fleqn,usenatbib]{mnras}

\usepackage{newtxtext,newtxmath}

\usepackage[T1]{fontenc}

\DeclareRobustCommand{\VAN}[3]{#2}
\let\VANthebibliography\thebibliography
\def\thebibliography{\DeclareRobustCommand{\VAN}[3]{##3}\VANthebibliography}

\usepackage{graphicx}	
\usepackage{amsmath}	
\usepackage{CJK}
\usepackage{bm}

\newcommand{\hi}{\textrm{H\textsc{i}}}

\newcommand{\secref}[1]{\hyperref[#1]{Section~\ref*{#1}}}
\newcommand{\appref}[1]{\hyperref[#1]{Appendix~\ref*{#1}}}

\title[Quadratic estimator for PCA]{A quadratic estimator view of the transfer function correction in intensity mapping surveys}

\author[Zhaoting Chen]{
Zhaoting Chen (陈兆庭) $^{1}$\thanks{E-mail: zhaoting chen@roe.ac.uk}
\\
$^{1}$Institute for Astronomy, The University of Edinburgh, Royal Observatory, Edinburgh EH9 3HJ, UK\\
}

\date{Accepted XXX. Received YYY; in original form ZZZ}

\pubyear{\the\year{}}

\defcitealias{2025MNRAS.537.3632M}{MK25}
\defcitealias{2023MNRAS.523.2453C}{C23}
\begin{document}
\begin{CJK*}{UTF8}{gbsn}
\label{firstpage}
\pagerange{\pageref{firstpage}--\pageref{lastpage}}
\maketitle

\begin{abstract}
In single dish neutral hydrogen (\hi) intensity mapping, signal separation methods such as principal component analysis (PCA) are used to clean the astrophysical foregrounds.
PCA induces a signal loss in the estimated power spectrum, which can be corrected by a transfer function (TF).
By injecting mock signals of \hi\ into the data and performing the PCA cleaning, we can use the cleaned mock \hi\ signal to cross-correlate with the original mock, and estimate the signal loss as a TF, $\mathcal{T}(\bm{k})$.
As expected, a correction of $\mathcal{T}(\bm{k})^{-1}$ restores the cross-power between the \hi\ and optical galaxies.
However, contrary to intuition, the \hi\ autopower also requires a $\mathcal{T}(\bm{k})^{-1}$ correction, not $\mathcal{T}(\bm{k})^{-2}$.
The $\mathcal{T}(\bm{k})^{-1}$ correction is only known empirically through simulations.
In this Letter, we show that the $\mathcal{T}(\bm{k})^{-1}$ correction in autopower is universal, and can be analytically proven using the quadratic estimator formalism through window function normalization.
The normalization can also be used to determine the TF correction for any type of linear process.
Using the window function, we demonstrate that PCA induces mode-mixing in the power spectrum estimation, which may lead to biases in the model inference.
\end{abstract}




\section{Introduction}
Neutral hydrogen (\hi) intensity mapping (e.g. \citealt{2001JApA...22...21B,2004MNRAS.355.1339B,2008PhRvL.100i1303C}) is an emerging probe for the large-scale structure (LSS) of the Universe.
\hi\ resides mostly inside galaxies after cosmic reionization (e.g. \citealt{2018ApJ...866..135V}), and can be measured using the 21\,cm line.
Tremendous progress has been made towards detecting the \hi\ clustering in the post-reionization Universe (e.g. \citealt{2013ApJ...763L..20M,2013MNRAS.434L..46S,2018MNRAS.476.3382A,2022MNRAS.510.3495W}).
In particular, the recent measurement from the MeerKAT Large Area Synoptic Survey (MeerKLASS; \citealt{2016mks..confE..32S}) L-band deep-field data reported in \cite{2025MNRAS.537.3632M}, hereafter \citetalias{2025MNRAS.537.3632M}, produces a cross-correlation detection.

A key technique for enabling the measurements of the \hi\ clustering is blind foreground removal (e.g. \citealt{2014MNRAS.441.3271W}).
The foregrounds in the radio waveband consist mainly of Galactic synchrotron radiation (e.g. \citealt{2015MNRAS.451.4311R}) and extragalactic radio sources (e.g. \citealt{1998AJ....115.1693C}), which are orders of magnitude brighter than the \hi.
{The foregrounds are spectrally smooth and therefore can be distinguished from the \hi\ signal.}
{In reality, however, observational systematics such as calibration errors and chromatic instrument beam couple with the sky signal, which complicate the spectral structure of the data (see \citealt{2025arXiv250403908C} in the case of MeerKLASS) and call for blind signal separation methods}.
The most commonly used method for foreground removal is Principal Component Analysis (PCA).
Given an intensity mapping data cube {that is mean-centred in each frequency channel}, $m(\bm{l},f)$, where $\bm{l}$ is the sky coordinate and $f$ is the observing frequency, we can calculate the frequency-frequency covariance of the data,
\begin{equation}
    \mathbfss{C}_{f_1f_2} = {\sum_{i=1}^{N_\theta} \bigg[(w\, m)(\bm{l}_i,f_1)\times(w\, m)(\bm{l}_i,f_2)\bigg] }\Big/{\sum_{i=1}^{N_\theta} \bigg[w(\bm{l}_i,f_1)w(\bm{l}_i,f_2)\bigg] },
\end{equation}
where $i$ iterates over each pixel in the angular plane, $N_\theta$ is the total number of pixels, and $w(\bm{l}_i,f)$ is the weight of each cube voxel.

Using the frequency-frequency covariance, we can perform an eigendecomposition and retrieve the eigenvectors of $\mathbfss{C}_{f_1f_2}$, $[\bm{v}_1,\bm{v}_2,...\bm{v}_{N_f}]$, where $N_f$ is the total number of frequency channels.
The eigenvectors are ranked from the largest eigenvalue to the smallest.
Assuming that the first $N_{\rm fg}$ eigenmodes contain all the foreground emission, we can form a matrix $\mathbfss{R}^{\rm PCA}$ so that
\begin{equation}
    \mathbfss{R}^{\rm PCA} = \mathbfss{I} - \sum_{m=1}^{N_{\rm fg}} \bm{v}_m\bm{v}_m^{\rm T},
\label{eq:rpca}
\end{equation}
where $\mathbfss{I}$ denotes the identity matrix and $^{\rm T}$ denotes the transpose of a matrix.
The cleaned map can then be calculated by multiplying the PCA matrix on the map data vector so that
\begin{equation}
    m_{\rm clean} (l,m,f_i) = \sum_j \mathbfss{R}^{\rm PCA}_{ij}\,m(l,m,f_j),
\end{equation}
where $j$ loops over each frequency channel and $_{ij}$ denotes the element of a matrix at the $i^{\rm th}$ row and $j^{\rm th}$ column.

For a sufficiently large number of modes $N_{\rm fg}$, {foregrounds should be strongly suppressed, and} the residual map $m_{\rm clean}$ can be used to estimate the \hi\ power spectrum.
However, removing the eigenmodes from the map also introduces signal loss.
To quantify the signal loss, a signal injection-based transfer function (TF) approach has been proposed \citep{2015ApJ...815...51S}.
Given a mock signal $m_{\rm \hi}$, we can inject the signal into the original map data $m$, and {recalculate the PCA matrix to perform the} cleaning to get the residual $m_{\rm clean}$.
The TF correction can be calculated as \citep{2023MNRAS.523.2453C}
\begin{equation}
    \mathcal{T}_{\rm \hi}(\bm{k}) = \bigg\langle \frac{\mathcal{P}[m_{\rm clean},m_{\rm \hi}]}{\mathcal{P}[m_{\rm \hi},m_{\rm \hi}]}\bigg\rangle,
\label{eq:tfauto}
\end{equation}
where $\mathcal{P}[,]$ denotes the operation of estimating the power spectrum by cross-correlating two fields and $\langle\rangle$ denotes the ensemble average over a number of realizations of mock.
Similarly, given a \hi\ mock signal and a corresponding galaxy mock density field $m_{\rm g}$, the TF for the cross-power spectrum can be calculated as
\begin{equation}
    \mathcal{T}_{\rm \hi,g}(\bm{k}) = \bigg\langle \frac{\mathcal{P}[m_{\rm clean},m_{\rm g}]}{\mathcal{P}[m_{\rm \hi},m_{\rm g}]}\bigg\rangle.
\label{eq:tfcross}
\end{equation}

\cite{2023MNRAS.523.2453C}, hereafter \citetalias{2023MNRAS.523.2453C}, find surprising properties of the TF.
First, it is found that for both cross-power and autopower, a correction of $\mathcal{T}(\bm{k})^{-1}$, multiplied on the estimated power spectrum after PCA cleaning, restores the correct amplitude of the underlying signal.
The conclusion is intuitive for cross-power.
However, for autopower, it is expected that $\mathcal{T}(\bm{k})^{-2}$ is needed instead of $\mathcal{T}(\bm{k})^{-1}$, as the TF defined in \autoref{eq:tfauto} is a cross-correlation between the cleaned and original mock signal, whereas the autopower is the autocorrelation of the cleaned signal.
Second, it is found that if the autopower is calculated by cross-correlating two data sets {of the same \hi\ signal} with foregrounds that are perturbed by different systematics (see Fig 12 of \citetalias{2023MNRAS.523.2453C}), the signal correction is further complicated and deviates from both $\mathcal{T}(\bm{k})^{-2}$ and $\mathcal{T}(\bm{k})^{-1}$ corrections.

In this Letter, we review the TF correction through the quadratic estimator formalism \citep{1997PhRvD..55.5895T}.
We show that the signal loss correction can be understood as a window function normalization, and link the normalization to the TF analytically.
The analytical expression of the TF is used to prove the $\mathcal{T}(\bm{k})^{-1}$ correction in the autopower, with numerical validations.

\section{Quadratic estimator}
\label{sec:qe}

\subsection{Fourier convention}
We start with choosing a Fourier convention.
The Fourier transform of an arbitrary function in a 3D rectangular box, $f(\bm{x})$, and its corresponding inverse transform, can be written as
\begin{equation}
    \tilde{f}(\bm{k}) = \int \frac{{\rm d}^3\bm{x}}{V} \,{\rm e}^{-i\,\bm{k}\cdot \bm{x}} f(\bm{x}), \, f(\bm{x}) = V \int \frac{{\rm d}^3\bm{k}}{(2\pi)^3} \,{\rm e}^{i\,\bm{k}\cdot \bm{x}} \tilde{f}(\bm{k}).
\label{eq:fft}
\end{equation}
where $\bm{k}$ is the 3D wavenumber vector and $V$ is the volume of the box.

For a tracer of dark matter $t$, the tracer field in a 3D rectangular box can be expressed as a data vector $\bm{d}_t$, whose $i^{\rm th}$ element is the value of the $i^{\rm th}$ grid at $\bm{x}_i$.
The Fourier transform of the tracer field, $\tilde{\bm{d}}_t$, can be written as
\begin{equation}
    \tilde{\bm{d}}_t = \mathcal{F}\, \bm{d}_t,
\end{equation}
where $\mathcal{F}$ is the discrete Fourier transform (DFT) matrix.
By discretizing \autoref{eq:fft}, we can see that
\begin{equation}
    \big[\mathcal{F}\big]_{ij} = \frac{1}{N} \,{\rm e}^{-i\,\bm{k}_i\cdot \bm{x}_j},
\label{eq:dft}
\end{equation}
where $[]_{ij}$ denotes the element of a matrix at the $i^{\rm th}$ row and $j^{\rm th}$ column, and $N$ is the total number of {grid points} in the box.
Note that the factor of $1/N$ in \autoref{eq:dft} corresponds to the ``forward'' normalization and is a consequence of the Fourier convention specified in \autoref{eq:fft}. The inverse DFT matrix can be written as
\begin{equation}
    \big[\mathcal{F}^{-1}\big]_{ij} = \,{\rm e}^{i\,\bm{k}_i\cdot \bm{x}_j}.
\label{eq:idft}
\end{equation}
It is straightforward to see that 
\begin{equation}
    \mathcal{F}^{\dagger}\mathcal{F} = \frac{1}{N}\,\mathbfss{I},
\end{equation}
where $^\dagger$ denotes the conjugate transpose of a matrix and $\mathbfss{I}$ denotes an identity matrix.

For any vector $\bm{v}$, its Fourier transform is $\tilde{\bm{v}} = \mathcal{F}\bm{v}$.
For any matrix $\mathbfss{M}$, its Fourier transform can be defined as
\begin{equation}
    \widetilde{\mathbfss{M}} = \mathcal{F}\, \mathbfss{M}\, \mathcal{F}^{-1},
\end{equation}
so that for any matrix $\mathbfss{M}$ and vector $\bm{v}$, $\widetilde{\mathbfss{M}}\, \tilde{\bm{v}} = \mathcal{F}\,\mathbfss{M}\,\bm{v}$.

\subsection{Window function}
We now turn to the quadratic estimator of the power spectrum.
For two tracers of dark matter which we denote as $t=1$ and $t=2$, the cross-power spectrum of the tracers in the 3D $\bm{k}$-grids is defined as
\begin{equation}
    V \bigg| \big\langle\tilde{d}_{1}(\bm{k}_\alpha)\tilde{d}_{2}^*(\bm{k}_\beta)  \big\rangle \bigg|_{\rm Re} = \delta^3_{\rm D}(\bm{k}_\alpha-\bm{k}_\beta)p^{12}_\alpha,
\label{eq:p12}
\end{equation}
where $p^{12}_\alpha$ denotes the \emph{underlying} 3D cross-power spectrum at $\bm{k}_\alpha$, $\tilde{d}_{1,2}$ denotes the Fourier transform of the \emph{underlying} tracer density fields, $^*$ denotes the complex conjugate, $\langle \rangle$ denotes the ensemble average, $||_{\rm re}$ denotes the real part of a complex number and $\delta^3_{\rm D}$ denotes the 3D delta function.
Note that, the autopower spectrum is a special case of \autoref{eq:p12} when 1 and 2 refer to the same tracer.
\autoref{eq:p12} can be written in a compact form,
\begin{equation}
    \mathbfss{C}_{12} = \frac{V}{2}\big\langle \tilde{\bm{d}}_1  \tilde{\bm{d}}_2^\dagger+ \tilde{\bm{d}}_2\tilde{\bm{d}}_1^\dagger  \big\rangle = \sum_\alpha \mathbfss{X}_\alpha \,p^{12}_\alpha,
\label{eq:c12}
\end{equation}
where $\mathbfss{X}_\alpha$ is the selection matrix that is zero for all elements except at the $\alpha^{\rm th}$ diagonal element,
\begin{equation}
    [\mathbfss{X}_\alpha]_{ij} = \delta^{\rm K}_{ij}\delta^{\rm K}_{i\alpha},
\end{equation}
where $\delta^{\rm K}$ is the Kronecker delta.

The observed tracer density field goes through a chain of linear operations before Fourier transform and cross-correlation.
For example, the underlying brightness temperature field of 21\,cm line is convolved with the primary beam of the telescope, calibrated by {applying calibration solutions}, gridded onto a sky map, PCA cleaned, frequency tapered, and finally gridded onto a rectangular grid with inverse noise variance weighting.
Since all of these operations are linear, we can collapse them into a matrix $\mathbfss{R}$, and write the quadratic estimator of the power spectrum as 
\begin{equation}
    \hat{p}^{12}_\alpha = \bigg|\frac{1}{2}{\bm{d}}_1^\dagger\mathbfss{E}^{12}_\alpha\bm{d}_2\bigg|_{\rm Re},
\label{eq:p12est}
\end{equation}
where $\mathbfss{E}^{12}_\alpha$ is the estimator matrix that can be decomposed as
\begin{equation}
    \mathbfss{E}^{12}_\alpha = V\,\mathbfss{R}_1^\dagger\mathcal{F}^\dagger \mathbfss{X}_\alpha\,\mathcal{F}\,\mathbfss{R}_2.
\end{equation}
Recall that $\mathcal{F}\, \mathbfss{R}\,\bm{d} = \widetilde{\mathbfss{R}}\,\tilde{\bm{d}}$,
and \autoref{eq:p12est} can be expanded so that
\begin{equation}
\begin{split}
    \hat{p}^{12}_\alpha =   \frac{V}{2}\bigg|\tilde{\bm{d}}_1^\dagger\widetilde{\mathbfss{R}}_{1}^\dagger \mathbfss{X}_\alpha \widetilde{\mathbfss{R}}_2\tilde{\bm{d}}_2\bigg|_{\rm Re}
    = \frac{V}{2}\bigg|{\rm tr}\Big[\widetilde{\mathbfss{R}}_{1}^\dagger \mathbfss{X}_\alpha \widetilde{\mathbfss{R}}_2 \tilde{\bm{d}}_2\tilde{\bm{d}}_1^\dagger \Big]\bigg|_{\rm Re},
\end{split}
\end{equation}
where tr[] denotes the trace of a matrix.

Combined with \autoref{eq:c12}, the ensemble average of the power spectrum estimator can be calculated so that
\begin{equation}
\begin{split}
    \langle \hat{p}^{12}_\alpha \rangle = & \frac{1}{2} \bigg| {\rm tr}\Big[ \widetilde{\mathbfss{R}}_{1}^\dagger \mathbfss{X}_\alpha \widetilde{\mathbfss{R}}_2\mathbfss{C}_{12} \Big] \bigg|_{\rm Re} 
    =  \frac{1}{2}\sum_\beta \bigg|{\rm tr}\Big[ \widetilde{\mathbfss{R}}_{1}^\dagger \mathbfss{X}_\alpha \widetilde{\mathbfss{R}}_2\mathbfss{X}_\beta\Big]p_\beta^{12}\bigg|_{\rm Re} \\
    = & \sum_\beta\bigg|(\widetilde{\mathbfss{R}}_1)_{\alpha\beta}(\widetilde{\mathbfss{R}}_2)^*_{\alpha\beta}\bigg|_{\rm Re}\,p^{12}_\beta
    =  \sum_\beta \Big(\mathbfss{H}\big[\mathbfss{R}_1,\mathbfss{R}_2\big]\Big)_{\alpha\beta}\,p^{12}_\beta,
\label{eq:window}
\end{split}
\end{equation}
where we obtain the \emph{unnormalized} window function of the estimator $\mathbfss{H}$ {as a function of the operation matrices $\mathbfss{R}_1$ and $\mathbfss{R}_2$,}
\begin{equation}
    \Big(\mathbfss{H}\big[\mathbfss{R}_1,\mathbfss{R}_2\big]\Big)_{\alpha\beta} \equiv \mathbfss{H}^{\mathbfss{R}_1,\mathbfss{R}_2}_{\alpha\beta}  = \bigg|(\widetilde{\mathbfss{R}}_1)_{\alpha\beta}(\widetilde{\mathbfss{R}}_2)^*_{\alpha\beta}\bigg|_{\rm Re},
\label{eq:hab}
\end{equation}
{where, from now on, we use the notation of $\mathbfss{H}^{\mathbfss{R}_1,\mathbfss{R}_2}$ to indicate the dependence of $\mathbfss{H}$ on the operation matrices for simplicity.}

\autoref{eq:window} states that, for a quadratic estimator of the power spectrum, the linear operations on the data vector leads to an effective linear operation on the power spectrum data vector, described by the window function $\mathbfss{H}$.
Therefore, one can normalize the window function by introducing a matrix $\mathbfss{M}$ so that
\begin{equation}
    \hat{p}'^{12}_\alpha = \sum_{\beta} \mathbfss{M}_{\alpha\beta}\, \hat{p}^{12}_\beta,
\end{equation}
which gives the final window function $\mathbfss{W} = \mathbfss{M}\,\mathbfss{H}$, with
\begin{equation}
    \langle\hat{p}'^{12}_\alpha \rangle =\sum_{\beta} \big(\mathbfss{M}\,\mathbfss{H}\big)_{\alpha\beta}\,{p}^{12}_\beta = \sum_{\beta} \mathbfss{W}_{\alpha\beta}\,p^{12}_\beta.
\end{equation}
One can choose $\mathbfss{M} = \mathbfss{H}^{-1}$ to decorrelate the mode-mixing or $\mathbfss{M} = \mathbfss{H}^{-1/2}$ to decorrelate the covariance of the estimator \citep{2002MNRAS.335..887T}.
In intensity mapping surveys, it is common to simply choose a diagonal matrix so that
\begin{equation}
    \mathbfss{M}_{\alpha\beta} = \delta^{\rm K}_{\alpha\beta} \Big(\sum_{\sigma}\mathbfss{H}_{\alpha\sigma}\Big)^{-1}
    = \mathcal{T}_{\rm true}^{-1}(\bm{k}_\alpha)\delta^{\rm K}_{\alpha\beta},
\label{eq:mab}
\end{equation}
so that the amplitude of the power spectrum is normalized, $\sum_\beta\mathbfss{W}_{\alpha\beta} = 1$.
Such a diagonal matrix is computationally efficient, since instead of performing a matrix multiplication, one can simply perform an element-wise multiplication on the estimated 3D $\bm{k}$-powers.
We define the required amplitude correction at each 3D $\bm{k}_\alpha$ mode as the true TF correction, $\mathcal{T}_{\rm true}^{-1}(\bm{k}_\alpha)$.

Most of the measurements of the \hi\ power spectrum in the literature implicitly uses this normalization.
For example, a weighting function including the frequency taper and inverse noise variance weighting, $w(\bm{x})$, can be multiplied to the map data.
The estimated power spectrum can then be normalized by rescaling the power spectrum by a factor of $N/(\sum_{\bm{x}}w^2(\bm{x}))$, which implicitly incorporates \autoref{eq:mab}.
The normalization also has a desired property, 
\begin{equation}
    \mathcal{T}_{\rm true}(\bm{k}_\alpha) = \sum_\sigma
    \bigg|(\widetilde{\mathbfss{R}}_1)_{\alpha\sigma}(\widetilde{\mathbfss{R}}_2)^*_{\alpha\sigma}\bigg|_{\rm Re} = \bigg|\Big(\widetilde{\mathbfss{R}}_1 \widetilde{\mathbfss{R}}_2^\dagger  \Big)_{\alpha \alpha} \bigg|_{\rm Re},
\label{eq:simple}
\end{equation}
which will be useful for the transfer function calculation, as we show in the following \secref{subsec:tf}.

\subsection{Transfer function}
\label{subsec:tf}
We now discuss the relation between the TF correction and the power spectrum normalization in \autoref{eq:mab}.

In the case of cross-power spectrum, the TF correction is calculated by taking the PCA matrix $\mathbfss{R}^{\rm PCA}$, applying it onto the \hi\ data vector, $\bm{d}_1$, and cross-correlating with a mock galaxy overdensity field, $\bm{d}_2$.
In the case of autopower spectrum, the TF correction is calculated by again taking the PCA matrix, applying it onto \hi\ data vector, $\bm{d}_1$, and cross-correlating with the \hi\ data vector itself.
In both cases, the TF term can be approximated as
\begin{equation}
\begin{split}
    \mathcal{T}(\bm{k}_\alpha) = & \bigg\langle \frac{\mathcal{P}[m_{\rm clean},m_{\rm g/\hi}]}{\mathcal{P}[m_{\rm \hi},m_{\rm g/\hi}]}\bigg\rangle \approx \frac{\sum_{\beta} \mathbfss{H}^{\mathbfss{R}^{\rm PCA},\mathbfss{I}}_{\alpha\beta}p^{12}_\beta}{ p^{12}_\alpha} \\ 
    = &\, \mathbfss{H}^{\mathbfss{R}^{\rm PCA},\mathbfss{I}}_{\alpha\alpha} = \Big|\widetilde{\mathbfss{R}}^{\rm PCA}_{\alpha\alpha}\Big|_{\rm Re} = \mathcal{T}_{\rm true}(\bm{k}_\alpha),
\end{split}
\label{eq:tfinh}
\end{equation}
where we use the fact that, when $\mathbfss{R}_1 = \mathbfss{R}^{\rm PCA}$ and $\mathbfss{R}_2 = \mathbfss{I}$, $\mathbfss{H}$ is a diagonal matrix. 
{Note that here the power spectrum $p^{12}_{\alpha}$ is the input power spectrum of the mock data vector.}
In previous work such as \citetalias{2023MNRAS.523.2453C}, the PCA matrix is re-calculated after mock injection so $\mathbfss{R}^{\rm PCA}$ in the TF calculation is different from the $\mathbfss{R}^{\rm PCA}$ used for the actual data.
The differences {in the resulting signal loss} are small, since the contribution of the mock \hi\ signal is small in the total frequency-frequency covariance.
This leads to the approximate equality in the above \autoref{eq:tfinh}.

We now compare \autoref{eq:tfinh} with the normalization described in \autoref{eq:mab}.
For cross-power, the galaxy overdensity is not PCA cleaned while the \hi\ is, giving the normalization \footnote{Note that, in the approximation of \autoref{eq:tfinh}, the TF terms for cross and autopower are the same, and we omit the subscripts in $\mathcal{T}_{\rm \hi}$ and $\mathcal{T}_{\rm \hi,g}$.}
\begin{equation}
    \mathbfss{M}^{\rm \hi,g}_{\alpha\beta} = \Big(\sum_{\sigma}\mathbfss{H}^{\mathbfss{R}^{\rm PCA,\mathbfss{I}}}_{\alpha\sigma}\Big)^{-1} \delta^{\rm K}_{\alpha\beta} \approx  \mathcal{T}(\bm{k}_\alpha)^{-1} \delta^{\rm K}_{\alpha\beta}. 
\end{equation}
As one can see, the required correction to normalize the window function is indeed $\mathcal{T}(\bm{k})^{-1}$ for the cross-power spectrum.
It also provides evidence to the fact that the TF correction is robust against the choice of mock signal modelling, since it does not depend on the mock data vectors.

Now we turn to the autopower spectrum, where the normalization according to \autoref{eq:mab} gives
\begin{equation}
    \mathbfss{M}^{\rm \hi}_{\alpha\beta} = \Big(\sum_{\sigma}\mathbfss{H}^{\mathbfss{R}^{\rm PCA},\mathbfss{R}^{\rm PCA}}_{\alpha\sigma}\Big)^{-1} \delta^{\rm K}_{\alpha\beta} ,
\label{eq:mauto}
\end{equation}
where $\mathbfss{H}^{\mathbfss{R}^{\rm PCA},\mathbfss{R}^{\rm PCA}}$ denotes the unnormalized window function when $\mathbfss{R}_1 = \mathbfss{R}_2 = \mathbfss{R}^{\rm PCA}$.

We now explore the explicit expression of $\sum_{\sigma}\mathbfss{H}^{\mathbfss{R}^{\rm PCA},\mathbfss{R}^{\rm PCA}}_{\alpha\sigma}$ in \autoref{eq:mauto} and its relation to $\sum_{\sigma}\mathbfss{H}^{\mathbfss{R}^{\rm PCA},\mathbfss{I}}_{\alpha\sigma}$.
The TF can be expanded by utilising the decomposition of the PCA matrix described in \autoref{eq:rpca}.
We first note that the Fourier transform of $\mathbfss{R}^{\rm PCA}$ can be expressed as
\begin{equation}
    \widetilde{\mathbfss{R}}^{\rm PCA} = \mathbfss{I} - \sum_m\mathcal{F}\,\bm{v}_m\,\bm{v}_m^\dagger\,\mathcal{F}^{-1} = \mathbfss{I} - N\sum_m \tilde{\bm{v}}_m\,\tilde{\bm{v}}^\dagger_m,
\end{equation}
where we use the fact that the Fourier transform of $\mathbfss{I}$ is itself and $\mathcal{F}^{-1} = N \mathcal{F}^{\dagger}$.
Using \autoref{eq:simple}, the TF in cross-power can then be calculated as
\begin{equation}
    \sum_{\sigma}\mathbfss{H}^{\mathbfss{R}^{\rm PCA},\mathbfss{I}}_{\alpha\sigma} 
    = \bigg|\widetilde{\mathbfss{R}}^{\rm PCA}_{\alpha\alpha}\bigg|_{\rm Re}
    =  1- N\sum_m\big|(\tilde{\bm{v}}_m)_\alpha\big|^2.
\label{eq:tfinv}
\end{equation}

Before we proceed to calculate $\mathbfss{H}^{\mathbfss{R}^{\rm PCA},\mathbfss{R}^{\rm PCA}}$, we comment on the physical meaning of \autoref{eq:tfinv}.
The eigenvectors of a matrix, as well as their Fourier transforms, are unitary and orthogonal, so that
\begin{equation}
    \forall \,[m_1,m_2],\ \bm{v}^{\rm T}_{m_1}\,\bm{v}_{m_2} = \delta^{\rm K}_{m_1m_2},\, \tilde{\bm{v}}^{\dagger}_{m_1}\,\tilde{\bm{v}}_{m_2} = \frac{\delta^{\rm K}_{m_1m_2}}{N},
\label{eq:fv1v2}
\end{equation}
where the extra factor of $1/N$ in the Fourier transform is due to the Fourier convention.
Using the orthogonality and unitarity of the eigenvectors, we can see that for any $\bm{k}$-mode $\bm{k}_\alpha$,
\begin{equation}
    0<\sum_{\sigma}\mathbfss{H}^{\mathbfss{R}^{\rm PCA},\mathbfss{I}}_{\alpha\sigma} = 1- N\sum_m\big|(\tilde{\bm{v}}_m)_\alpha\big|^2 < 1.
\end{equation}
Therefore, the signal loss of PCA originates from the fact that, in the power spectrum estimation, a fraction of the power spectrum amplitude is projected out by the estimator.
The level of signal loss is determined by the length of the eigenvectors in Fourier space at $\bm{k}_\alpha$.
For example, for the foreground dominated modes, we expect a smooth eigenvector $\bm{v}$, which corresponds to a Fourier vector that has a peak at small $k_\parallel$.
As a result, the signal loss is most severe at the $\bm{k}$-modes corresponding to large physical scales along the line-of-sight.

Furthermore, one can see that
\begin{equation}
\begin{split}
    \widetilde{\mathbfss{R}}^{\rm PCA}& \big(\widetilde{\mathbfss{R}}^{\rm PCA}\big)^\dagger =  \Big(\mathbfss{I} -2 N\sum_{m} \tilde{\bm{v}}_{m}\,\tilde{\bm{v}}^\dagger_{m} + N^2 \sum_{m_1,m_2} \tilde{\bm{v}}_{m_1}\,\tilde{\bm{v}}^\dagger_{m_1}\tilde{\bm{v}}_{m_2}\,\tilde{\bm{v}}^\dagger_{m_2} \Big)\\
    = & \,\mathbfss{I} - N\sum_{m} \tilde{\bm{v}}_{m}\,\tilde{\bm{v}}^\dagger_{m} = \widetilde{\mathbfss{R}}^{\rm PCA},
\end{split}
\label{eq:rsqure}
\end{equation}
where we utilize the orthogonality in \autoref{eq:fv1v2}.
Combining \autoref{eq:rsqure} and \autoref{eq:simple}, we can calculate the auto TF to be
\begin{equation}
\begin{split}
    \sum_{\sigma}&\mathbfss{H}^{\mathbfss{R}^{\rm PCA},\mathbfss{R}^{\rm PCA}}_{\alpha\sigma} 
    = \bigg|\Big(\widetilde{\mathbfss{R}}^{\rm PCA} \big( \widetilde{\mathbfss{R}}^{\rm PCA}\big)^\dagger  \Big)_{\alpha \alpha} \bigg|_{\rm Re} = \bigg|\Big(\widetilde{\mathbfss{R}}^{\rm PCA}  \Big)_{\alpha \alpha} \bigg|_{\rm Re}\\ 
    = & \sum_{\sigma}\mathbfss{H}^{\mathbfss{R}^{\rm PCA},\mathbfss{I}}_{\alpha\sigma} .
\end{split}
\label{eq:hautoinv}
\end{equation}

The corresponding normalization is therefore
\begin{equation}
    \mathbfss{M}^{\rm \hi}_{\alpha\beta} = \Big(\sum_{\sigma}\mathbfss{H}^{\mathbfss{R}^{\rm PCA},\mathbfss{R}^{\rm PCA}}_{\alpha\sigma}\Big)^{-1} \delta^{\rm K}_{\alpha\beta}  \approx \mathcal{T}(\bm{k}_\alpha)^{-1} \delta^{\rm K}_{\alpha\beta}.
\end{equation}

\autoref{eq:hautoinv} shows that, regardless of the numerical details of the PCA, the attenuation of the \hi\ autopower spectrum amplitude is the same as the attenuation in the \hi-galaxy cross-power, which can both be approximated by the TF, $\mathcal{T}$.
As a result, in both auto and cross, the required signal loss correction is $\mathcal{T}(\bm{k})^{-1}$.
From the derivation of \autoref{eq:hautoinv}, one can see that the $\mathcal{T}(\bm{k})^{-1}$ correction stems from the fact that the eigenvectors are orthogonal to each other.
The derivations provide a formal proof to the phenomenon seen in the simulations in \citetalias{2023MNRAS.523.2453C} and the discussions in Appendix B3 of \citetalias{2025MNRAS.537.3632M}.

The derivation also provides insights into the internal cross-correlation where both data sets are \hi\ signals but possess different foregrounds.
The $\mathcal{T}(\bm{k})^{-1}$ correction derived in this work for autopower is specific to the quadratic estimator, with $\mathbfss{R}_1 = \mathbfss{R}_2 = \mathbfss{R}^{\rm PCA}$.
For two data sets with different PCA cleaning, the orthogonality no longer holds, and therefore the signal loss correction is no longer $\mathcal{T}(\bm{k})^{-1}$.
The quadratic estimator formalism, on the other hand, is universal.
For any given $\mathbfss{R}_1$ and $\mathbfss{R}_2$, the window function can be calculated exactly using \autoref{eq:hab} (see e.g. \citealt{2021MNRAS.501.1463K} for the case of using Gaussian Process Regression). It can also be used to mitigate the systematics in the data (see e.g. \citealt{2021JCAP...07..016W,2022PhRvD.106d3534W,2024arXiv240808949W}).

We note an important fact that, while the TF corrects the amplitude by normalising over $\sum_\sigma \mathbfss{H}_{\alpha\sigma}$, it does not account for the mode-mixing in the window function of the autopower.
If there is severe contamination that requires aggressive cleaning, the TF correction may not be robust as the window function mixes a wide range of scales at small $|\bm{k}|$.
The mode-mixing also affects the calculation of covariance, which will be important for model inference in future data analysis.

\section{Numerical Validation}
\label{sec:num}

In this section, we verify the analytical derivations in \secref{sec:qe} with numerical simulations, and use the simulations to demonstrate the effects of TF on the power spectrum estimation.
For simplicity, we neglect all observational effects and simply generate \hi\ temperature fields in comoving boxes to isolate the effects of PCA.
The box dimensions and \hi\ model parameters are listed in \autoref{tab:spec}, which corresponds to the survey volume of \citetalias{2025MNRAS.537.3632M}.
Using the model \hi\ power spectrum, we generate Gaussian realizations of the \hi\ density field.
A synchrotron emission cube is also generated using the Global Sky Model \citep{2017MNRAS.464.3486Z} under the flat-sky approximation.
The density field combining the \hi\ and the synchrotron radiation is then used for PCA cleaning.
We chose $N_{\rm fg} = 40$, which leads to extreme signal loss and mode-mixing at large scales, to demonstrate the window function $\mathbfss{H}$ induced by the cleaning.
{For comparison, the MeerKLASS 2019 L-band data requires $N_{\rm fg} = 4$ on large scales \citep{2024arXiv241206750C}, and the 2021 data described in \citetalias{2025MNRAS.537.3632M} requires $N_{\rm fg} = 10$.} 
Note that, in the flat sky approximation, the PCA cleaning is only operating on the data vector along the $z$-axis, and thus the window function is only non-trivial along the $k_\parallel$ direction.

\begin{table}
    \centering
    \begin{tabular}{c|c|c|c|c}
        $L_{x,y,z}\,$(Mpc) & $N_{x,y,z}$ & $z$ & $\Omega_{\hi}$ & $b_{\hi}$ \\\hline
        $[636,\,424,\,263]$ & [72,\,48,\,252] & 0.42 & $0.5\times10^{-3}$ & 1.5 \\\hline
    \end{tabular}
    \caption{The specifications for the \hi\ simulations. $L_{x,y,z}$ are the lengths of the box in each dimension, $N_{x,y,z}$ are the numbers of grids in each dimension, $z$ is the redshift, $\Omega_{\hi}$ is the \hi\ density relative to the critical density of the present day, and $b_{\hi}$ is the \hi\ bias. The specifications are chosen to match the survey in \citetalias{2025MNRAS.537.3632M}. We also assume the cosmology reported in \citet{2020A&A...641A...6P}.}
    \label{tab:spec}
\end{table}

The resulting 3D power spectrum is then averaged into cylindrical {and 1D} $\bm{k}$-space for visualization.
{For the cylindrical power spectrum,} we choose the $|\bm{k}_\perp|$ bins to be linearly spaced between [0,\,0.5]$\,\rm Mpc^{-1}$ with 10 bins, and the $|k_\parallel|$ bins to be linearly spaced between [0,\,3.0]$\,\rm Mpc^{-1}$ with 15 bins.
{For the 1D power spectrum, we choose the $|\bm{k}|$ bins to be logarithmically spaced between [0.005,\,3.0]$\,\rm Mpc^{-1}$ with 10 bins.}
The simulations are repeated with 100 independent realizations of the \hi\ signal and averaged over the realizations for demonstration.

\begin{figure}
    \centering
    \includegraphics[width=1.0\linewidth]{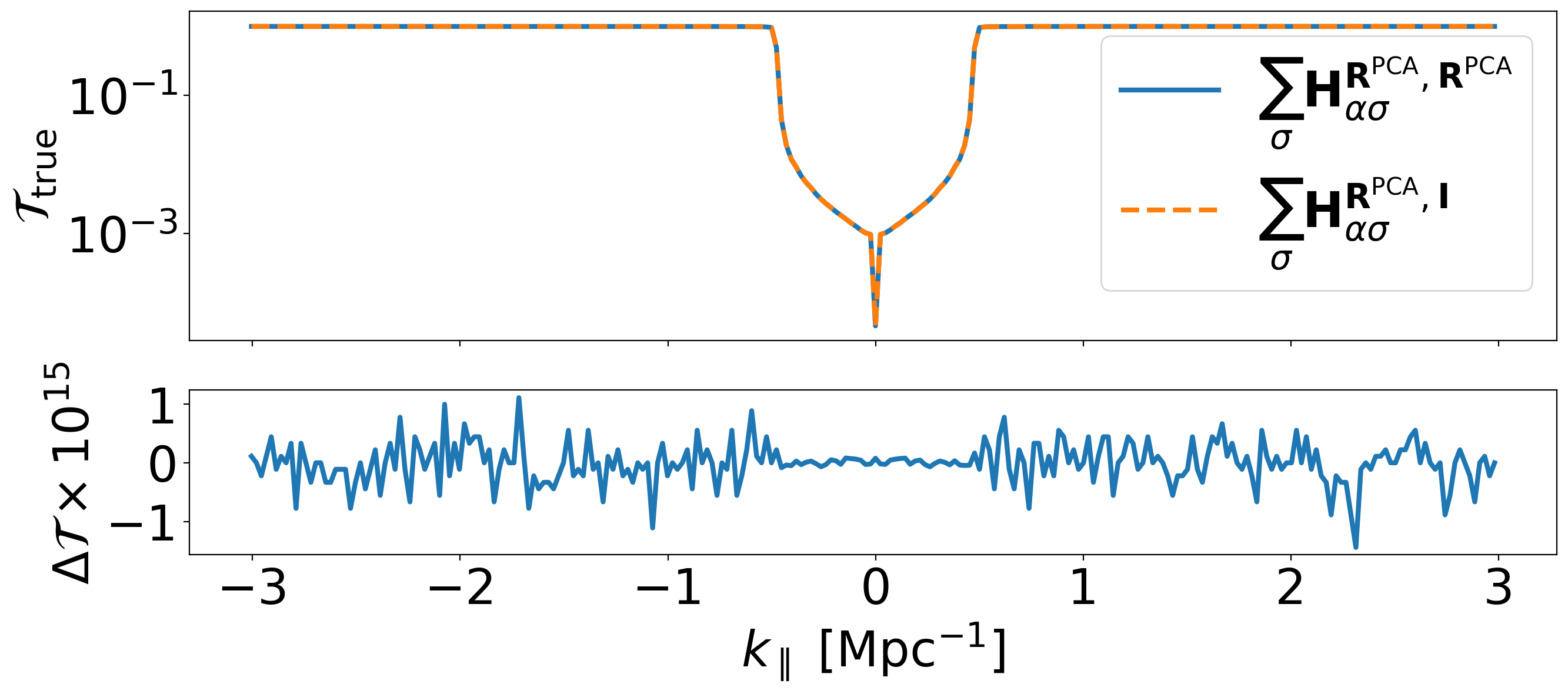}
    \caption{The upper panel shows the transfer function based on the window function normalization, $\sum_{\sigma} \mathbfss{H}_{\alpha\sigma}$, for the \hi\ autopower spectrum ($\sum_{\sigma} \mathbfss{H}_{\alpha\sigma}^{\mathbfss{R}^{\rm PCA},\mathbfss{R}^{\rm PCA}}$) and \hi-galaxy cross-power spectrum ($\sum_{\sigma} \mathbfss{H}_{\alpha\sigma}^{\mathbfss{R}^{\rm PCA},\mathbfss{I}}$). The lower panel shows the $\lesssim 10^{-15}$ differences between the transfer functions for auto and cross-power, which are numerical precision artefacts. For illustration, the signal loss is chosen to be extreme, with $N_{\rm fg} = 40$, and does not represent a realistic case in data analysis.}
    \label{fig:DeltaT}
\end{figure}

\begin{figure*}
    \centering
    \includegraphics[width=0.49\linewidth]{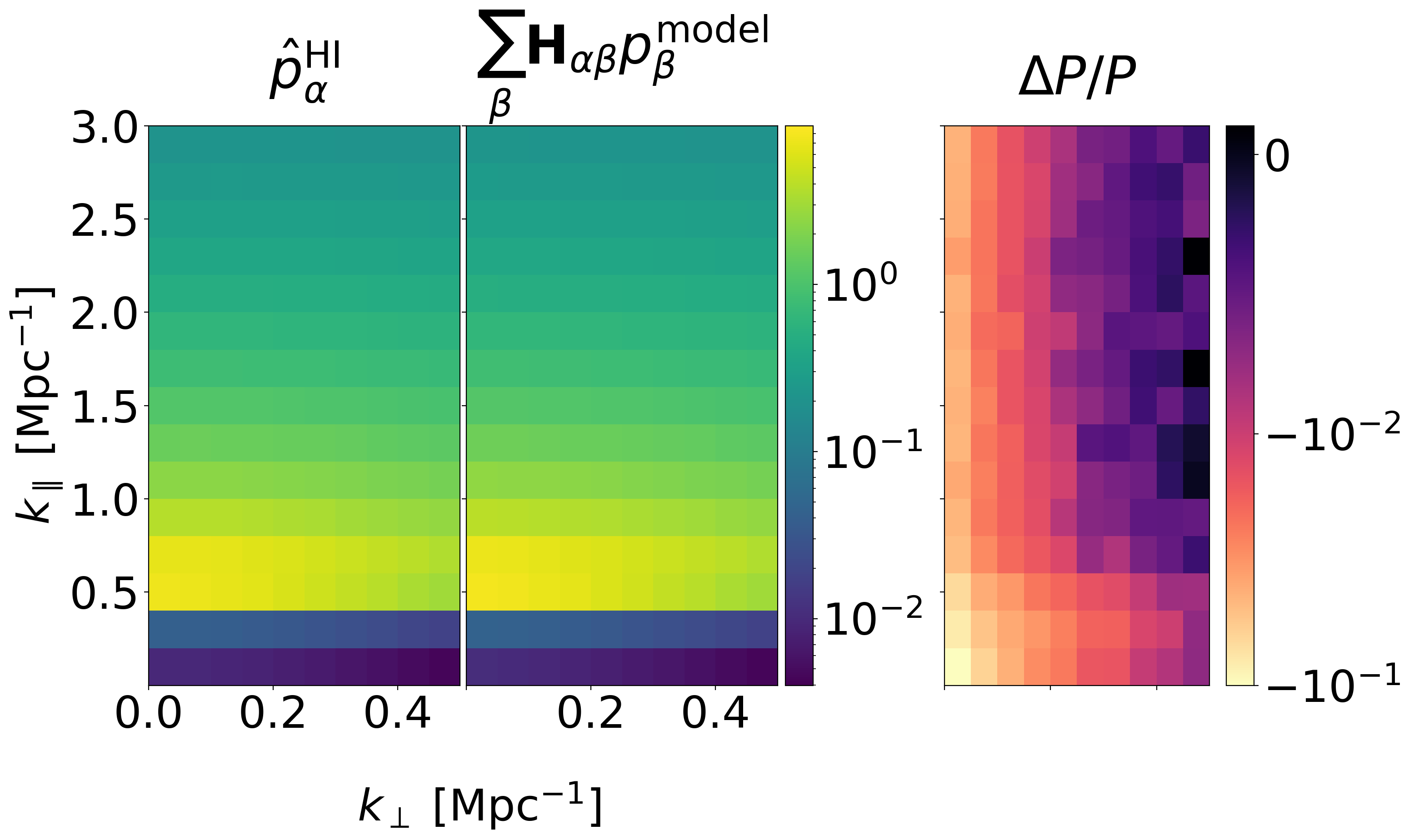}
    \includegraphics[width=0.49\linewidth]{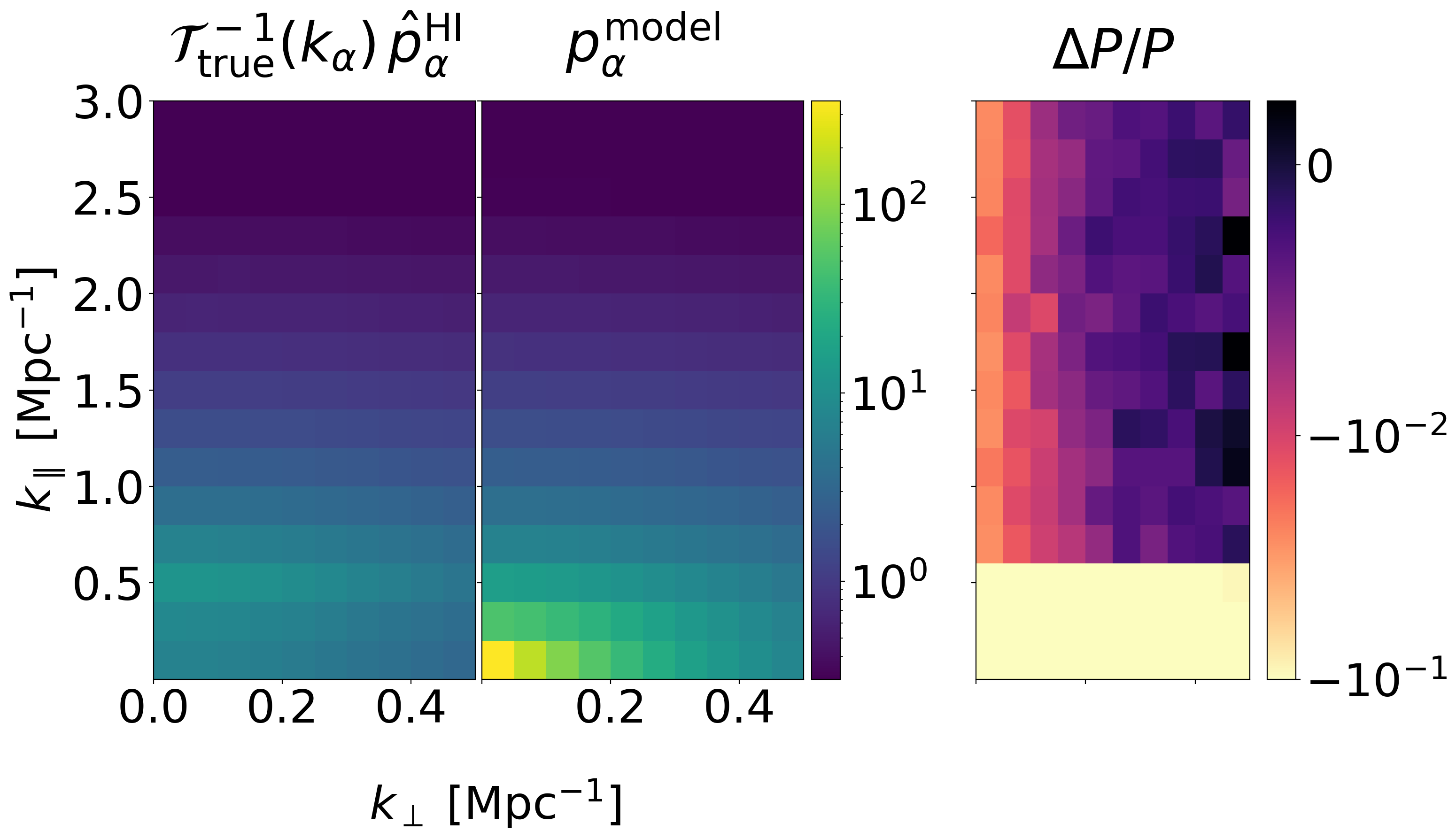}
    \caption{
    The left half of the panels show the output \hi\ power spectrum from simulations after PCA cleaning (``$\hat{p}^{\rm HI}_{\alpha}$''), the input model \hi\ power spectrum with window function $\mathbfss{H}$ applied (``$\sum_{\beta}\mathbfss{H}_{\alpha\beta}\,{p}^{\rm model}_{\beta}$''), and the fractional differences between the former and the latter (``$\Delta P/P$'').
    The right half of the panels show the output \hi\ power spectrum after PCA cleaning with TF corrections applied (``$\mathcal{T}(k_\alpha)^{-1}\,\hat{p}^{\rm HI}_{\alpha}$''), the input \hi\ power spectrum (``${p}^{\rm model}_{\alpha}$''), and the fractional differences between the two (``$\Delta P/P$''). All power spectra are in the unit of $\rm mK^2\,Mpc^3$. In the fractional differences, values below -0.1 are set to -0.1 for visualization.
    For illustration, the signal loss is chosen to be extreme, with $N_{\rm fg} = 40$, and does not represent a realistic case in data analysis.
    }
    \label{fig:pscy}
\end{figure*}

We first verify \autoref{eq:hautoinv} by numerically calculating $\sum_{\sigma}\mathbfss{H}^{\mathbfss{R}^{\rm PCA},\mathbfss{R}^{\rm PCA}}_{\alpha\sigma}$ and $\sum_{\sigma}\mathbfss{H}^{\mathbfss{R}^{\rm PCA},\mathbfss{I}}_{\alpha\sigma}$ to show that they are equal to each other.
The results are shown in \autoref{fig:DeltaT}.
As expected, the signal loss, quantified by the TF, is severe at small $|k_\parallel|$, due to the foregrounds being smooth.
In the extreme setting of removing $N_{\rm fg} = 40$ modes, the signal loss becomes significant at $|k_\parallel|\lesssim 0.5\,{\rm Mpc^{-1}}$.
The TFs are identical for auto and cross-power, with the numerical differences $<10^{-15}$.
The results again verify the fact that the TF correction in autopower is $\mathcal{T}^{-1}$, not $\mathcal{T}^{-2}$.

We then verify that the \hi\ power spectrum after PCA cleaning indeed follows the window function $\mathbfss{H}$.
For comparison, we calculate the 3D power spectrum of the PCA-cleaned \hi\ field, $\hat{p}^{\rm HI}_{\alpha}$, and average it into cylindrical $\bm{k}$-space.
The PCA cleaning gives the cleaning matrix $\mathbfss{R}^{\rm PCA}$, which is used to calculate the window function $\mathbfss{H}$.
We then apply the window function to the input model spectrum, $\sum_{\beta}\mathbfss{H}_{\alpha\beta}\,{p}^{\rm model}_{\beta}$, which is then averaged into the cylindrical $\bm{k}$-space as well for comparison.
The results are shown in the left half of \autoref{fig:pscy}.
As seen, the output \hi\ power spectrum, after cleaning, suffers severe signal loss at $k_\parallel \lesssim 0.5\,{\rm Mpc^{-1}}$, consistent with the TF corrections shown in \autoref{fig:DeltaT}.
Moreover, the model power spectrum, after applying the window function, is almost identical with the output \hi\ power spectrum, with the differences $\lesssim 10\%$ even with extreme $\sim 10^{-3}$ signal loss leading to large numerical uncertainties.
It validates the window function expression described in \autoref{eq:hab}.

With our extreme setting of removing $N_{\rm fg}=40$ modes, we can explore the limitation of the TF correction which we show in the right half of \autoref{fig:pscy}.
The TF correction, $\mathcal{T}^{-1}$, is applied to the output \hi\ power spectrum and compared against the input model power spectrum.
At $k_\parallel \lesssim 0.5\,{\rm Mpc^{-1}}$ where signal loss is severe, the correction does not match the input model.
This is because, while the TF ensures that the final window function obeys $\sum_\sigma \mathbfss{W}_{\alpha\sigma} = 1$, the measured power at small $k_\parallel$ is a mixture of different $k_\parallel$ scales, with large $k_\parallel$ modes having much smaller amplitude.

\begin{figure}
    \centering
    \includegraphics[width=0.7\linewidth]{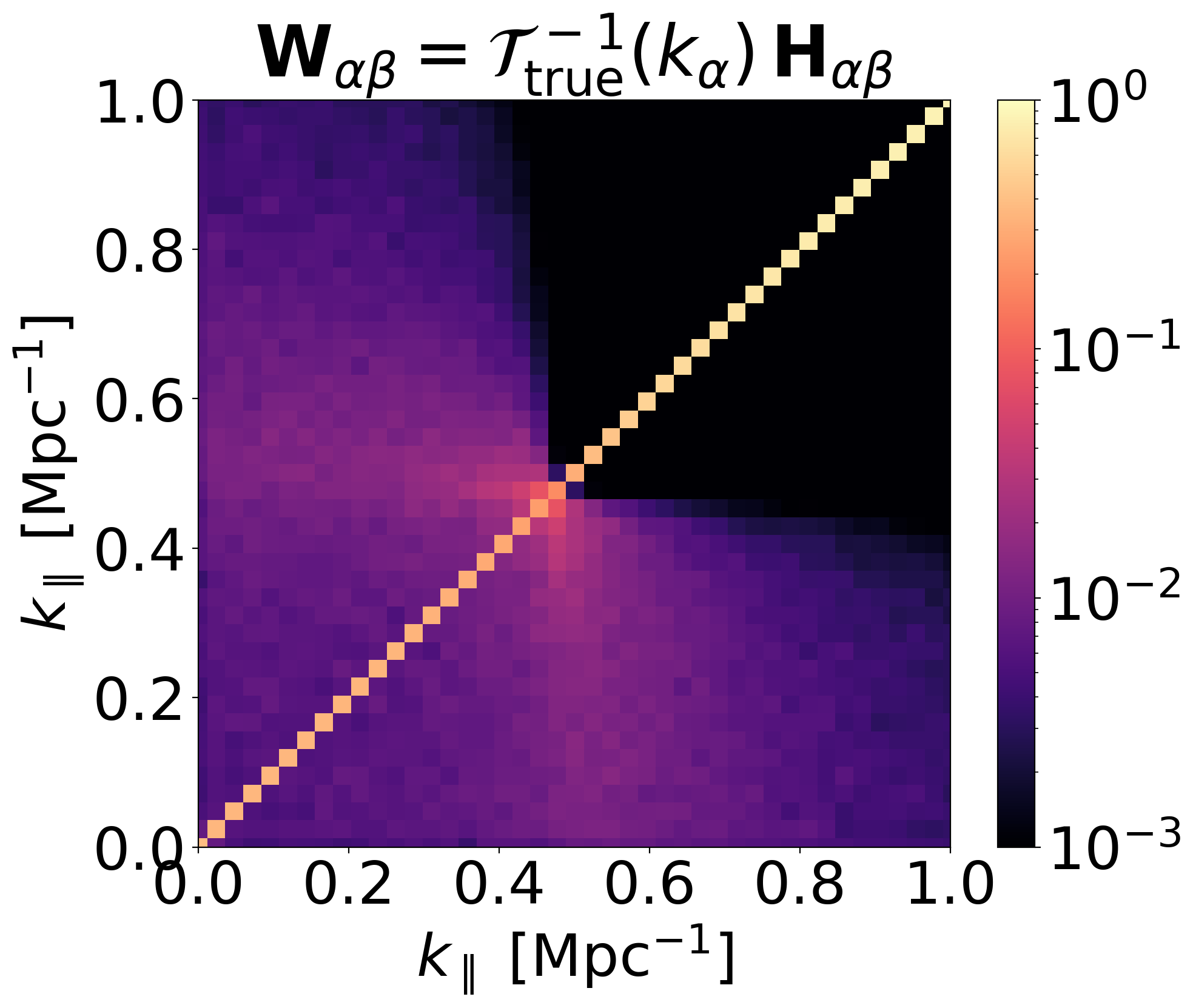}
    \caption{The normalized window function $\mathbfss{W} = \mathbfss{M}\,\mathbfss{H}$ described in \autoref{eq:window}, with the normalization matrix $\mathbfss{M}$ following \autoref{eq:mab}. Values below $10^{-3}$ are set to $10^{-3}$ for visualization. The $k_\parallel >1\,{\rm Mpc^{-1}}$ range is omitted to highlight the mode-mixing at large scales.
    For illustration, the signal loss is chosen to be extreme, with $N_{\rm fg} = 40$, and does not represent a realistic case in data analysis.
    }
    \label{fig:window}
\end{figure}

\begin{figure}
    \centering
    \includegraphics[width=1.0\linewidth]{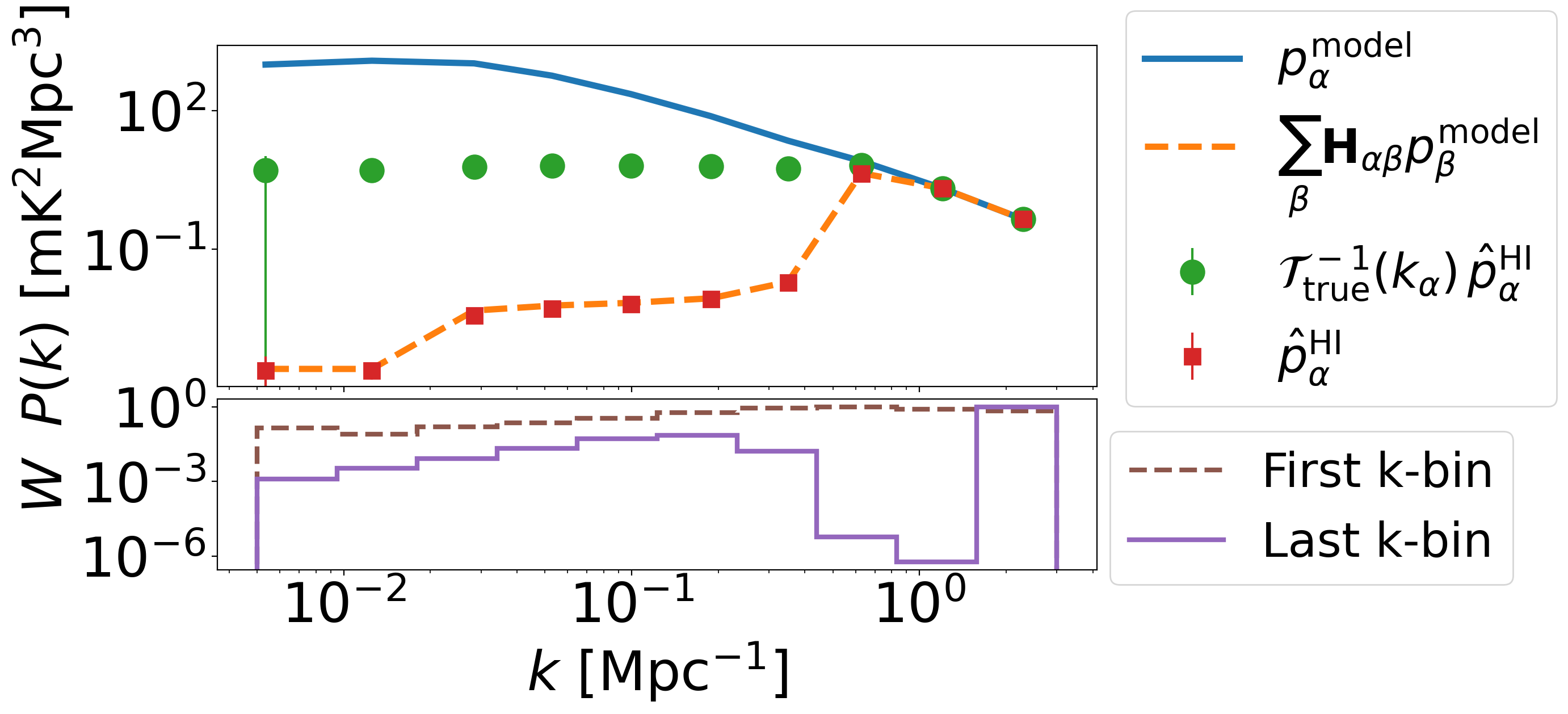}
    \caption{{Upper panel: The 1D average of the power spectra (see the caption of \autoref{fig:pscy}). The error bar corresponds to the standard deviation among the realizations. Lower panel: The normalized 1D window function for the first and the last $|\bm{k}|$ bin.}
    }
    \label{fig:1d}
\end{figure}

The mode-mixing is visualized in \autoref{fig:window}, where the normalized window function $\mathbfss{W}$ is shown.
Corresponding to the signal loss, there is severe mode-mixing at $k_\parallel \lesssim 0.5\,{\rm Mpc^{-1}}$.
Measurements at these scales contain a large contribution from $k_\parallel \gtrsim 0.5\,{\rm Mpc^{-1}}$ modes, which have smaller power spectrum amplitude.
As a result, even though the correct TF correction is applied, the corrected power spectrum does not match the input.
In current data analysis, the signal loss is far less severe compared to the case in this work, and the region of $\bm{k}-$space with significant mode-mixing is much smaller than the $k_\parallel \lesssim 0.5\,{\rm Mpc^{-1}}$ shown here.
However, failure to consider the mode-mixing will still lead to biased estimation of the power spectrum and its covariance.
{To further illustrate this, we compute the 1D power spectrum and the 1D window function in \autoref{fig:1d}. The input power spectrum with window function applied (``$\sum_{\beta}\mathbfss{H}_{\alpha\beta}\,{p}^{\rm model}_{\beta}$'') agrees well with the output (``$\hat{p}^{\rm HI}_{\alpha}$''). Meanwhile, the transfer function corrected output power (``$\mathcal{T}(k_\alpha)^{-1}\,\hat{p}^{\rm HI}_{\alpha}$'') deviates significantly from the input (``${p}^{\rm model}_{\alpha}$''), and is almost constant at $|\bm{k}|\lesssim 0.5\,{\rm Mpc}^{-1}$. As seen in the lower panel, the measurement at small $|\bm{k}|$ contains a mixture of all scales and very little information from the desired $|\bm{k}|$ bin, leading to the constant power in $\mathcal{T}(k_\alpha)^{-1}\,\hat{p}^{\rm HI}_{\alpha}$. While mode-mixing is not severe at large $|\bm{k}|$, it can be seen that the measurement contains a mixture of small $|\bm{k}|$. The biasing has also been seen in the data as described in Section 7 of \citet{2024arXiv241206750C}.}
In the future, where the \hi\ power spectrum is detected at scales larger than the Baryon Acoustic Oscillations, the biases will have a significant impact on cosmological inference.

We note that, the numerical TF correction studied in \citetalias{2023MNRAS.523.2453C} is not exactly equivalent to the TF correction shown in this section.
In data analysis, the \hi\ signal is obtained as sky maps and gridded onto rectangular grids.
The effects of PCA cleaning are then no longer only along the $k_\parallel$-axis. 
The corresponding $\mathbfss{R}^{\rm PCA}$ and window function $\mathbfss{H}$ are much larger in their dimensions and harder to compute numerically.
The TF calculation based on mock injection provides an efficient way of estimating the window function amplitude.
On the other hand, since PCA cleaning is not the last operation in the overall $\mathbfss{R}$, the operations of weighting and gridding may further complicate the window function and break the assumption in \autoref{eq:tfinh}.
Further development is needed to apply the formalism shown here to the data analysis, which we leave for future work.

{Finally, we emphasize that the quadratic estimator formalism in this work provides insights into improving the TF correction. The approximation in \autoref{eq:tfinh} can be dropped if we redefine the TF correction to be}
\begin{equation}
    \mathcal{T}'(\bm{k}_\alpha) =   \frac{\big\langle\mathcal{P}[m'_{\rm clean},m_{\rm g/\hi}]\big\rangle}{\big\langle\mathcal{P}[m_{\rm \hi},m_{\rm g/\hi}]\big\rangle} = \mathcal{T}_{\rm true}(\bm{k}_\alpha),
\end{equation}
{where $m'_{\rm clean}$ is the mock \hi\ signal cleaned by the PCA matrix obtained from the data without the injection.
The alternative TF correction is, in fact, studied in the upper panel of Figure C1 of  \citetalias{2023MNRAS.523.2453C} and found to be underestimating the \hi\ power.
The results shown in this work demonstrate that this underestimation is due to the mode-mixing.
In other words, the default TF correction in \citetalias{2023MNRAS.523.2453C} treats the mode-mixing \emph{as if it were part of the signal loss}.
Follow-up work should instead use the alternative TF correction which is completely model independent, and expand the TF formalism to account for the mode-mixing separately.
}

\section{Conclusion}
\label{sec:conclusion}
In this Letter, we have shown that the transfer function correction in \hi\ intensity mapping experiments can be understood through the quadratic estimator formalism.
The TF correction is effectively a normalization of the power spectrum estimator, which restores the amplitude of the measured power spectrum.
The true correction depends only on the foreground removal operator, and therefore the TF correction can be robust against the choice of mock signal modelling.
Using the quadratic estimator formalism, we prove analytically that the TF correction should be $\mathcal{T}(\bm{k})^{-1}$ for both auto and cross-power.
The quadratic estimator formalism, {with the derivations up to \autoref{eq:mauto}}, is universal {for all linear operations on the data vector} and can be used for more complicated scenarios, for example for cross-correlating two \hi\ data sets.

The quadratic estimator formalism also shows that the TF correction ignores mode-mixing in the autopower caused by foreground removal.
The formalism outlined in this work can be used to further develop the TF technique to include the mode-mixing, which will be the focus of the follow-up work.
Constructing an accurate window function will be crucial for enabling cosmological inference using \hi\ intensity mapping in the era of the SKAO.

\vspace{-0.1cm}

\section*{Acknowledgements}
ZC thanks Alkistis Pourtsidou, Steven Cunnington, Stefano Camera, Isabella Paola Carucci, Jos\'e Luis Bernal, Matilde Barberi-Squarotti for discussions, and Zheng Zhang for pointing out a way to simplify the derivation.
ZC is funded by a UKRI Future Leaders Fellowship [grant MR/X005399/1; PI: Alkistis Pourtsidou].

\section*{Data Availability}
The code for generating the simulation results in this Letter can be found \href{https://github.com/zhaotingchen/qe_tf_im/tree/main}{here}.



\bibliographystyle{mnras}
\bibliography{example} 




\appendix


\bsp	
\label{lastpage}
\end{CJK*}
\end{document}